\documentclass{jfm}
\usepackage{amsmath}
\usepackage{amsfonts}
\usepackage{amssymb}
\usepackage[utf8x]{inputenc}
\usepackage{graphicx}
\usepackage{natbib}
\usepackage{float}
\usepackage{comment}
\usepackage{epstopdf}
\usepackage{lineno}
\usepackage{scalerel,stackengine}
\usepackage{color}

\title{Third-order structure functions for isotropic turbulence
  with bidirectional energy transfer} \author{Jin-Han Xie and
  Oliver B\"uhler} \affiliation{Courant Institute of Mathematical
  Sciences, New York University, New York, NY 10012, USA }

\newcommand{\br}[1]{\left( #1 \right)}
\newcommand{\sbr}[1]{\left[ #1 \right]}

\newcommand{\ex}{\mathrm{e}}
\newcommand{\ii}{\mathrm{i}}

\newcommand{\mc}[1]{\mathcal{#1}}

\newcommand{\bu}{\boldsymbol{u}}
\newcommand{\bv}{\boldsymbol{v}}

\newcommand{\bB}{\boldsymbol{B}}
\newcommand{\bx}{\boldsymbol{x}}

\newcommand{\bF}{\boldsymbol{F}}

\newcommand{\bk}{\boldsymbol{k}}

\newcommand{\bV}{\boldsymbol{V}}

\newcommand{\bor}{\boldsymbol{r}}

\newcommand{\ovl}[1]{\overline{#1}}

\newcommand{\dd}{\mathrm{d}}

\newcommand\reallywidehat[1]{\arraycolsep=0pt\relax%
	\begin{array}{c}
		\stretchto{
			\scaleto{
				\scalerel*[\widthof{\ensuremath{#1}}]{\kern-.5pt\bigwedge\kern-.5pt}
				{\rule[-\textheight/2]{1ex}{\textheight}} %WIDTH-LIMITED BIG WEDGE
			}{\textheight} % 
		}{0.5ex}\\           % THIS SQUEEZES THE WEDGE TO 0.5ex HEIGHT
		#1\\                 % THIS STACKS THE WEDGE ATOP THE ARGUMENT
		\rule{-1ex}{0ex}
	\end{array}
}

\begin{document}
\maketitle

\begin{abstract}
	
%	The Kolmogorov 4/5-law for third-order structure functions in the inertial range of homogeneous isotropic turbulence is a cornerstone of modern turbulence study.
  We derive and test a new heuristic theory for third-order
  structure functions that resolve the forcing scale in the
  scenario of simultaneous spectral energy transfer to both small
  and large scales, which can occur naturally in rotating
  stratified turbulence or magnetohydrodynamical~(MHD)
  turbulence, for example.  The theory has three parameters,
  namely the upscale/downscale energy transfer rates and the
  forcing scale, and it includes the classic inertial range
  theories as local limits.  When applied to measured data, our
  global-in-scale theory can deduce the energy transfer rates
  using the full range of data, therefore it has broader
  applications compared with the local theories, especially in
  the situations where the data is imperfect.  In addition,
  because of the resolution of forcing scales, the new theory can
  detect the scales of energy input, which was impossible before.
  We test our new theory with a two-dimensional simulation of MHD
  turbulence.
	
\end{abstract}

\section{Introduction}

The direction of spectral energy transfer is a crucial feature of
a turbulent system. In contrast to isotropic two- and
three-dimensional turbulence, where the spectral energy transfer
is predominantly  to large or small scales
\citep{Kraichnan1967,Kolmogorov1941}, some systems such as
rotating stratified turbulence and magnetohydrodynamical
turbulence \cite[cf.][]{Alexakis2018} exhibit more complex
behavior, where the energy transfer can be
bidirectional\footnote{In \cite{Alexakis2018} this scenario is 
  termed a ``split energy cascade'', but we think ``bidirectional
  energy transfer'' is more intuitive.}.

To quantify the magnitude and the direction of
energy transfer, theories that link the measurable third-order
structure functions to energy fluxes have been developed in the
inertial ranges, which are away from both the dissipation and
forcing scales for isotropic turbulent systems.  E.g., in
three-dimensional (3D) isotropic turbulence, where energy transfers
downscale, \citet{Kolmogorov1941} found that the longitudinal
third-order structure function $\ovl{\delta u_L^3}$ and the energy
input rate $\epsilon$, which equals the magnitude of energy flux
in a statistically steady state, are exactly related by
$\ovl{\delta u_L^3}=-\frac{4}{5}\epsilon r<0$, where $r$ is the
distance between the two measured points in the inertial range.
In contrast, energy transfers to large scales in two-dimensional (2D)
turbulence, and the corresponding relation in the energy inertial
range becomes $\ovl{\delta u_L^3}=\frac{3}{2}\epsilon r>0$
\citep{Bernard1999,Lindborg1999,Yakhot1999}.  It has since become
commonplace to use local fits to power laws of observed
third-order structure functions to detect spectral energy
transfer directions in a variety of systems
\citep{Lindborg2007,Kurien2006,Deusebio2014}, including the solar
wind \citep{Sorriso-Valvo2007} and atmospheric flow
\citep{Cho2001}.  %ob added the following
Indeed, sometimes just the \textsl{sign} of the observed third-order
structure function has been used to estimate the direction of the
energy flux at some scale $r$.  This is not a robust diagnostic
once we consider the shortcomings of the local theories.

First, 
they are valid only in local inertial
ranges, which are far away from both the forcing and dissipation
scales. Thus, when applying to measured data, one has to
determine where the inertial range is in the first place,  i.e., in
order to use the \citet{Kolmogorov1941}
theory one needs to find where the third-order structure
function is linear in $r$.  But it is possible that
different researchers choose different data ranges, and inertial
ranges might be  short and hard to
identify using imperfect measured data, all of which leads to
uncertainties.   Second, the local, inertial-range theories
by definition fail at the forcing scales, which prevents
the important detection of   forcing scales, e.g., for 
geophysical flows.  Third, previous theories were developed for
scenarios with unidirectional energy transfer, but
there is good evidence that in natural turbulence, e.g.,
in the atmosphere and oceans, energy transfers simultaneously to both large and small
scales  \citep{Marino2015,Pouquet2017}.  The
direction of energy flux is essential for these
structure-function theories, e.g., \citet{Xie2018} illustrate how
the 3D \citep{Kolmogorov1941} and 2D
\citep{Kraichnan1967} turbulence must be treated differently when
taking the infinite Reynolds number limit in the
K\'arm\'an-Howarth-Monin (KHM) \cite[cf.][]{Monin1975,Frisch1995}
equation, because of their opposite directions of energy
transfer. So it is questionable to directly apply the previous
theories to scenarios with bidirectional energy transfer.

Thus, we want to obtain a forcing-resolving global-in-scale theory
that not only captures different inertial ranges in one formula
but also applies to bidirectional energy transfer, allowing us to
make use of the measured data over a wide range that includes
the forcing scales.  Here, we derive such a theory for
isotropic 2D turbulence and test it against a numerical simulation of
2D MHD turbulence with bidirectional energy transfer
\citep{Seshasayanan2014,Seshasayanan2016}, which is a limiting
case of 3D MHD with strong background magnetic field
\citep{Gallet2015}.  We also show how to adapt our theory to
turbulent flows in 1D or 3D.

\section{Theoretical framework}

We start from the generic K\'arm\'an-Howarth-Monin (KHM) equation
for two-point correlations, in which the nonlinear terms appear
via the divergence of a third-order vector field:
\begin{equation}
\frac{1}{2}\frac{\partial}{\partial t}C - \frac{1}{4}\nabla\!\cdot\!\bV = D + P. \label{abs_eq}
\end{equation}
Here $C$ is the second-order correlation function, $\bV$ is a
vector of third-order structure functions if the system has a
quadratic nonlinearity, and $D$ and $P$ describe the effects of
dissipation and external forcing, respectively.  For example, in
the case of 2D homogeneous isotropic turbulence studied by \cite{Xie2018},
\begin{subequations}
	\begin{align}
	C &= \ovl{\bu\!\cdot\!\bu'},\\
	\bV &= \ovl{\delta\bu |\delta\bu|^2},\\
	D &= -\alpha\ovl{\bu\!\cdot\!\bu'}+\nu\nabla^2\ovl{\bu\!\cdot\!\bu'}, \label{2d_damp}\\
	P &= \frac{1}{2}\br{ \ovl{\bF\!\cdot\!\bu'} + \ovl{\bF'\!\cdot\!\bu} },
	\end{align}
\end{subequations}
where $\bu'=\bu(\bx+\bor)$ with $\bor$ the displacement between
two measurement points, $\delta\bu=\bu'-\bu$, $\alpha$ is a
Rayleigh damping rate, $\nu$ is the viscosity, $\bF$ is the
external forcing and the overline denotes the ensemble average.
For statistically steady turbulent states, (\ref{abs_eq})
simplifies to
\begin{equation}
-\frac{1}{4}\nabla\!\cdot\!\bV = D + P \label{abs_eq_stea}.
\end{equation} 
The Fourier transform of $C$ yields the power spectrum as a
function of wavenumber $\bk$ so by
applying the Fourier transform to (\ref{abs_eq}) and integrating
over the wavenumber shell $|\bk| < K$ it follows that 
(cf.\ \S 6 in \cite{Frisch1995})
\begin{equation}
F(K) = -\int_{|\bk|\leq K}^{} \frac{1}{4}\widehat{\nabla\!\cdot\!\bV} \dd \bk \label{F_1}
\end{equation}
is the nonlinear spectral energy transfer rate across the
wavenumber shell with radius $K$, i.e., a positive $F(K)>0$
measures the downscale
energy transfer from larger scales ($|\bk| < K$) to smaller scales
($|\bk| > K$) in spectral space.  Under the assumption of
isotropy, the third-order structure-function vector is
\begin{equation}
\bV = V(r) \hat{\bor},
\end{equation}
where $r=|\bor|$ and $\hat{\bor}$ is a unit vector pointing in the direction of $\bor$. 
Thus, in two dimensions (\ref{F_1}) can be expressed as (cf.\ (5.8) in \cite{Xie2018}) 
\begin{equation}
F(K) = -\frac{K^2}{4}\int_{0}^{\infty}  V(r) J_2(Kr) \dd r, \label{F_V}
\end{equation}
where $J_2$ is the second-order Bessel function.
Equivalently, using the orthogonality of Bessel functions, we can invert (\ref{F_V}) to obtain
\begin{equation}
V(r) = -4r\int_{0}^{\infty}\frac{1}{K}F(K)J_2(Kr) \dd K. \label{V_F}
\end{equation}

\subsection{Non-dissipative theory}
\label{sec:non-diss-energy}

\begin{figure}
	\centering
	\includegraphics[width=0.99\linewidth]{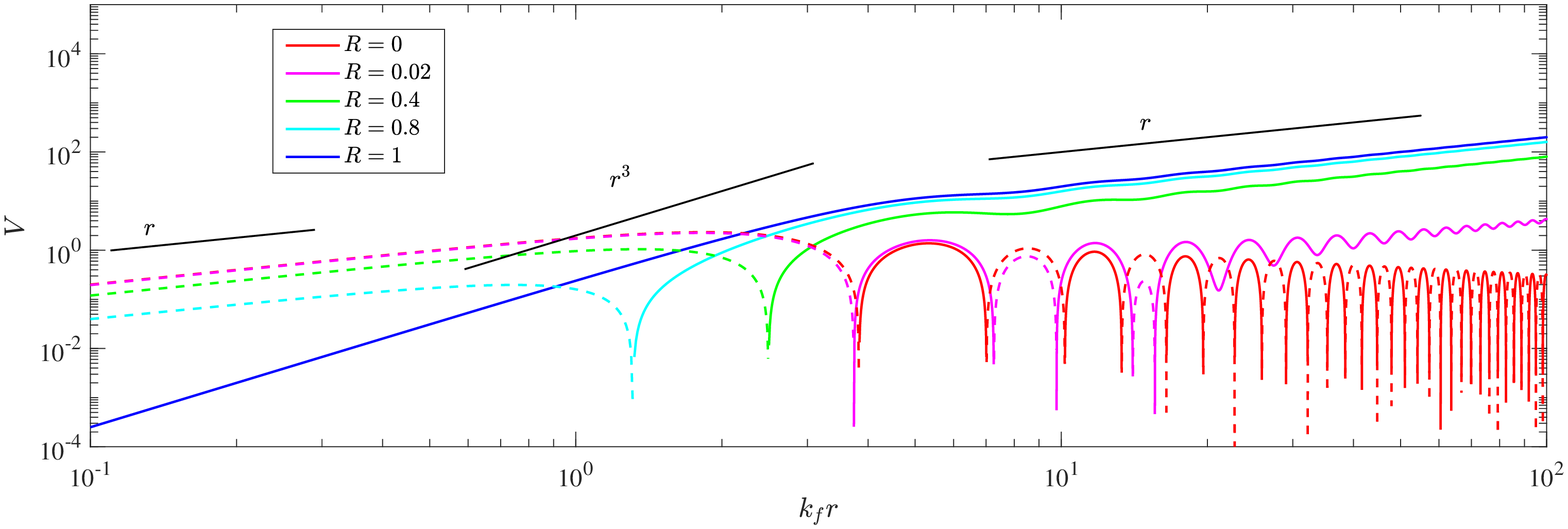}
	\caption{Theoretical expression (\ref{V_theo}) with
          $k_f=1$, $\epsilon=1$ and different values of
          $R\equiv\epsilon_\mathrm{u}/\epsilon$. At the right end
          ($k_fr=10^2$), the curves align with descending $R$
          from above to below.  Solid and dashed lines denote
          positive and negative values, respectively. The black
          lines illustrate classical power laws.}
	\label{fig_V_illus}
\end{figure}

We now consider first an idealized non-dissipative scenario where the external forcing is
sharply localized at some wavenumber $k_f$  with corresponding length scale
$l_f=1/k_f$ whilst the dissipation at small and large scales
has been pushed to $K\to \infty$ and $K\to 0$,
respectively.    Corrections due to finite-scale dissipation are deferred
until \S~\ref{simba}.  Therefore,
at finite $K$ we can argue that  $F(K)$ must take the piecewise
constant form
\begin{equation}
F(K) =-\epsilon_\mathrm{u} + (\epsilon_\mathrm{u}+\epsilon_\mathrm{d}) H(K-k_f). \label{F_heavi}
\end{equation}
Here $\epsilon_\mathrm{u} $ and $\epsilon_\mathrm{d}$ are the
magnitudes of upscale and downscale energy fluxes, $H$ is the
Heaviside function, and
\begin{equation}
  \label{eq:1}
  \epsilon=\epsilon_\mathrm{u}+\epsilon_\mathrm{d}
\end{equation}
is the total
energy input rate.
Substituting (\ref{F_heavi}) into (\ref{V_F}) yields the
corresponding non-dissipative expression 
\begin{equation} 
V(r) = 2\epsilon_\mathrm{u} r - 4 \frac{\epsilon}{k_f}J_1(k_fr), \label{V_theo}
\end{equation}
%From the perspective of dimensional analysis, the third-order
%structure function should follow
%$V=C(k_fr,\epsilon_\mathrm{u}/\epsilon_\mathrm{d})\epsilon_\mathrm{d}r$.
which extends Kolmogorov's classical inertial range theory by including the
forcing scale as well as bidirectional energy transfer.
In contrast to the classic definition of
inertial range, we do not assume that the considered scale $r$ is
far away from the forcing scale $l_f$, thus, (\ref{V_theo}) is a
forcing-scale-resolving expression and we call it a global
solution.  We illustrate its behavior in Figure~\ref{fig_V_illus}
using $k_f=1$,
$\epsilon=1$, and various values of the fractional upscale flux
\begin{equation}
  \label{eq:2}
  R = \frac{\epsilon_\mathrm{u}}{\epsilon}.
\end{equation}
In the limit $R=1$ of completely upscale energy flux the present (\ref{V_theo})
reduces to the second equality in (4.9) already derived in
\cite{Xie2018}.
%ob added last bit
Notably, $V(r)$ is sign-definite and positive \textsl{only} if $R=1$,
i.e., for all values $R<1$ the sign of $V(r)$ changes at least once.
%ob moved this block
Also, the case with $R=0.02$ is almost indistinguishable from the
limiting case $R=0$ for downward-only energy flux, but only if
$k_fr\ll 10$.  Otherwise their difference becomes obvious as
$k_fr\gg10$.  In the intermediate range ($k_fr\sim10$), where $r$
is larger than the forcing scale $1/k_f$ and almost all energy
transfers upscale, the structure function $V$ with $R=0.02$ still
has alternating signs, which illustrates once more that one
cannot safely read off the direction of the spectral transfer just from
the sign of the third-order structure function.

Naturally, in the limits of large and small $k_fr$ the
global expression (\ref{V_theo}) recovers the classic local
results \cite[cf.][]{Bernard1999,Lindborg1999,Yakhot1999}
asymptotically, i.e.,
\begin{equation}\label{V_expan}
V(r) =\left\{\begin{matrix}
\underset{\mathrm{downscale\,energy}}{\underbrace{-2\epsilon_\mathrm{d} r}} + \underset{\mathrm{``enstrophy"}}{\underbrace{\dfrac{1}{4}\epsilon k_f^2r^3}} + O\br{(k_fr)^5}, \quad \mathrm{when} \quad k_fr\ll 1, \\
\underset{\mathrm{upscale\,energy}}{\underbrace{2\epsilon_u r}} + O\br{(k_fr)^{-1/2}}, \quad \mathrm{when} \quad k_fr\gg 1.
\end{matrix}\right.
\end{equation}
Interestingly, the small-scale ``enstrophy" term recovers the classical enstrophy
cascade result of 2D
turbulence when $\epsilon_\mathrm{d}=0$, but if
$\epsilon_\mathrm{d}\neq0$ there may be not even be any 
enstrophy conservation in the turbulent system, but nonetheless
this term arises in all cases in the expansion of $V(r)$.

\subsection{Dissipative corrections}
\label{simba}

For realistic turbulence dissipation brings about corrections to
(\ref{V_theo}) at large and small $r$.  E.g., in 2D turbulence a
linear Ekman damping introduces the log-correlation to the energy
spectrum at the enstrophy inertial range \citep{Kraichnan1971},
and it bounds the range of inverse energy cascade
\cite[e.g.][]{SmithKS2002}.  We do not want to introduce a
closure that links second- and third-order structure functions to
calculate the shape of them, instead, we simply derive an exact
relation that links them diagnostically.  The derivation starts
from distinguishing the large- and small-scale damping terms
which dominantly absorb upscale and downscale energy fluxes,
respectively.  This distinction is necessary because the two
types of damping influence the inertial range different in the
limit of zero viscosity: the large-scale damping brings about
leading-order contribution while the effect of small-scale
damping is of higher order compared with that of the external
forcing, as shown in \cite{Xie2018}.  Let's write the dissipation
term in (\ref{abs_eq}) as
\begin{equation}
  \label{eq:3}
  D=\mc{L}_{D} C= D_l + D_s = \mc{L}_{Dl} C + \mc{L}_{Ds}C 
\end{equation}
where the operator $\mc{L}_{D}$ is the sum of large- and
small-scale parts $\mc{L}_{Dl}$ and $\mc{L}_{Ds}$, respectively.
For example, \cite{Xie2018} used $\mc{L}_{Dl}=-\alpha$ and $\mc{L}_{Ds}=\nu\nabla^2$
for Rayleigh damping and Navier--Stokes diffusion.   The large- and
small-scale net dissipation rates are then $\epsilon_\mathrm{u} =D_l|_{r=0} $ and
$\epsilon_\mathrm{d}=D_s|_{r=0}$, respectively. 
For two-dimensional isotropic turbulence integrating
(\ref{abs_eq_stea}) over a disk of radius $r$ yields 
\begin{equation}\label{V_damp0}
%\begin{aligned}
V_{d}(r) %&= -\frac{4}{r}\int_{0}^{r}s\br{D(s)+P(s)}\dd s &\\ &
= -\frac{4}{r}\int_{0}^{r}sD_s(s)\dd s -\frac{4}{r}\int_{0}^{r}s\br{D_l(s)+\epsilon_\mathrm{u}}\dd s + 2\epsilon_\mathrm{u}r  -\frac{4}{r}\int_{0}^{r}sP(s)\dd s. %&
%\end{aligned}
\end{equation}
If the external forcing is white-noise in time and centered at
wavenumber $k_f$ then  (\ref{V_damp0}) becomes
\begin{equation}\label{V_damp1}
\begin{aligned}
V_{d} = -\frac{4}{r}\int_{0}^{r}sD_s(s)\dd s -\frac{4}{r}\int_{0}^{r}s\br{D_l(s)+\epsilon_\mathrm{u}}\dd s + 2\epsilon_\mathrm{u}r  -4\frac{\epsilon}{k_f}J_1(k_fr).
\end{aligned}
\end{equation}
This is the sought-after dissipative correction to (\ref{V_theo}).  We need to
note that for a general 2D turbulence system we are not able to
strictly derive that in the limit of zero viscosity the finite
damping effect tends to zero and is therefore negligible compared
with the limit result, and to do so we need to consider a
specific turbulence system with prescribed damping term, one such
example is 2D turbulence studied by \citet{Xie2018}.  Note that
in the derivation we need to distinguish large- and small-scale
dissipations.  But the smallness of the finite damping effect in
the zero viscosity limit matches the derivation starting from the
idealized spectral energy flux (\ref{F_heavi}).  In the next
section we check both the non-dissipative  result (\ref{V_theo}) and its
dissipation correction (\ref{V_damp1}) in a MHD example.

\section{Application to two-dimensional  MHD turbulence} \label{sec_MHD}

To test our heuristic theory we performed numerical simulations
of a 2D MHD turbulent flow in which the velocity $\bv$ and the
magnetic field $\bB$ are coplanar.  This is an ideal test system
because \citet{Seshasayanan2014} found bidirectional energy
transfer in this 2D system, and also its KHM equation has the
generic form (\ref{abs_eq}) with a third-order structure
function vector \cite[cf.][]{Podesta2008} defined by 
\begin{equation}
\bV = \ovl{\delta\bu \br{\delta\bu\!\cdot\!\delta\bu}}  + \ovl{\delta\bu \br{\delta\bB\!\cdot\!\delta\bB}} -2\ovl{\delta\bB \br{\delta\bB\!\cdot\!\delta\bu}}. \label{V_MHD}
\end{equation}
Here the magnetic field is normalized to have velocity units
such that $C= \ovl{\bu\!\cdot\!\bu'} + \ovl{\bB\!\cdot\!\bB'}$.

The numerical simulation uses a Fourier pseudospectral method
with 2/3 dealiasing in space, a resolution $512\times512$ and a
%ob
%
fourth-order explicit Runge--Kutta scheme in time, in which the
nonlinear terms are treated explicitly and linear terms
implicitly using an integrating factor method. We take the
forcing wavenumber to be $k_f=32$, the momentum and magnetic
equation are forced by random forces which are white-noise in
time, and we control the kinetic energy input rate to be 100
times of that of the magnetic energy, which is a case that is
found to have bidirectional energy transfer
\citep{Seshasayanan2016}. We add hypoviscosity with operator
$\nabla^{-2}$ and hyperviscosity with operator $\nabla^6$ to both
the velocity and magnetic fields to dissipate energy transferred
to large and small scales, respectively, and therefore the
turbulence system reaches a statistically steady state.

\begin{figure}
	\centering
	\includegraphics[width=0.49\linewidth]{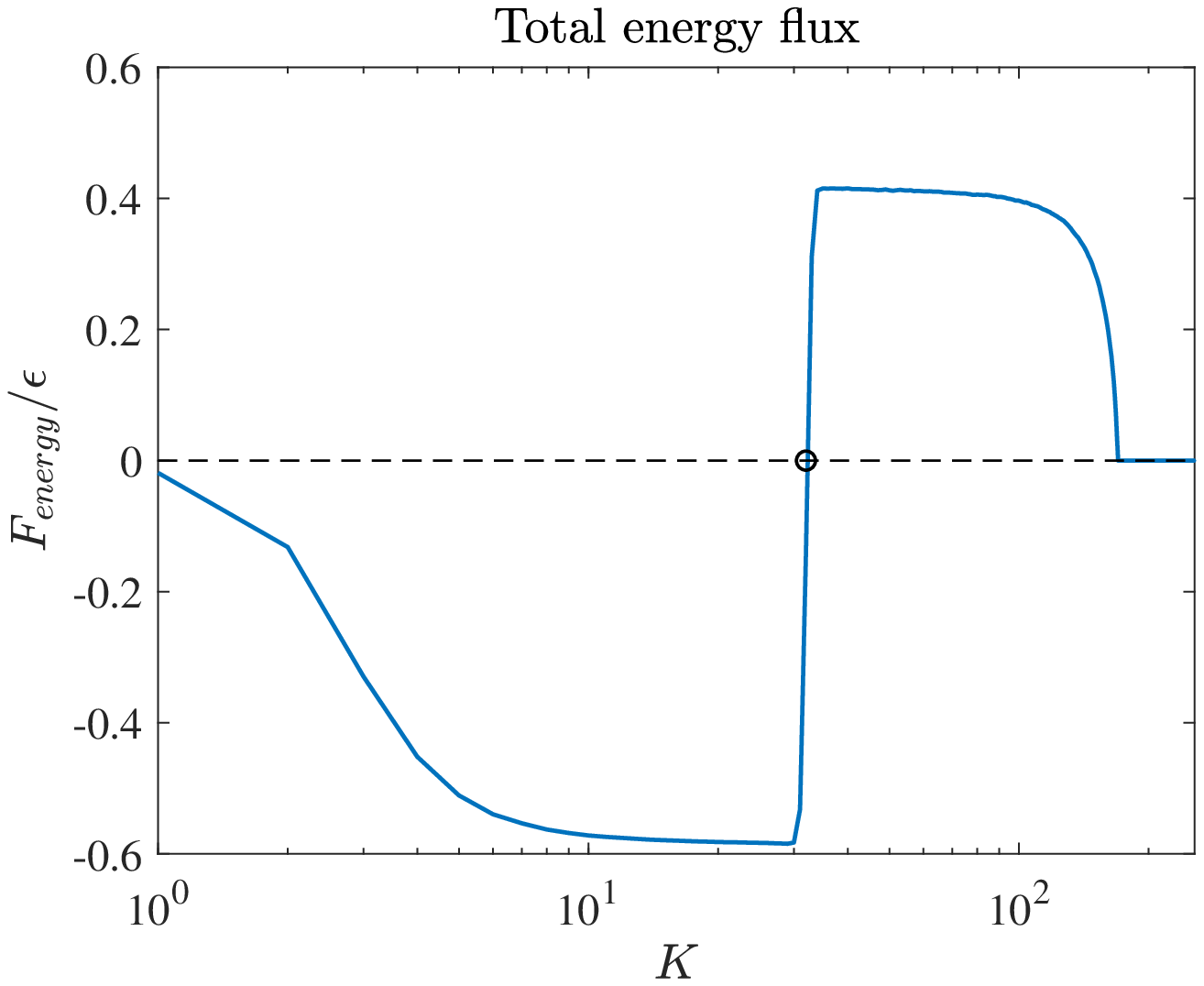}
	\includegraphics[width=0.49\linewidth]{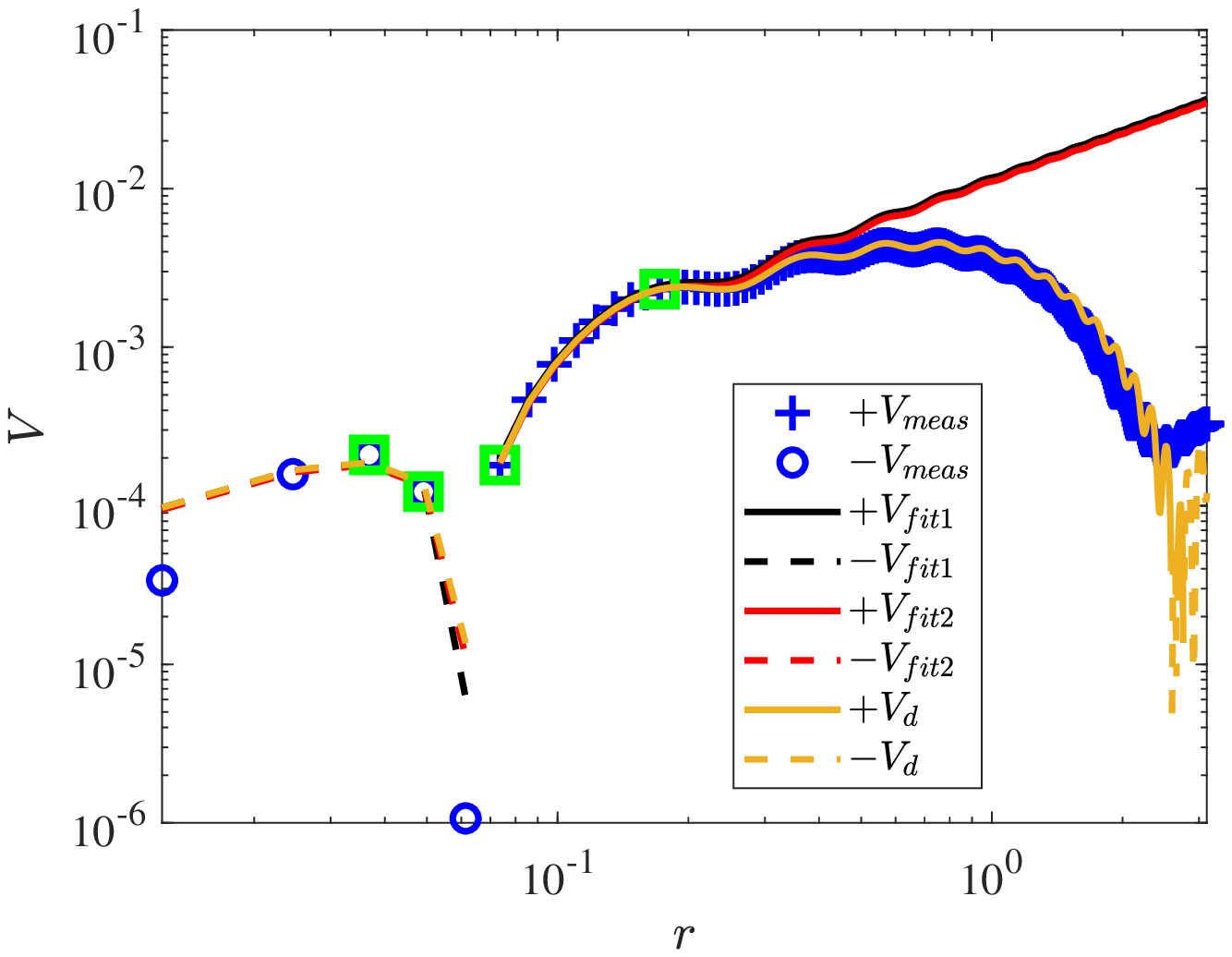}
	\caption{Left panel: total observed energy transfer
          normalized by the total energy input rate
          $\epsilon$. The circle marks the forcing wavenumber
          $k_f=32$. Right panel: Comparison of third-order
          structure functions obtained from the statistics of
          numerical data (blue), two zero-viscosity fitting
          curves (black and red), and the finite damping fitting
          curve (yellow). The four green boxes mark the four
          points used for fitting 2 (red). In the legend, the
          symbols ``$+$" and ``$-$" denote positive and
          negative values, respectively.} 
	\label{fig_energy_trans_V}
\end{figure}

We show in the left panel of Figure~\ref{fig_energy_trans_V} the
spectral transfer $F(K)$ of total energy, which is the sum of
kinetic and magnetic energy. Here the spectral energy transfer is
directly calculated in Fourier space from the pseudospectral code
without making use of the third-order structure function in
physical space.  As expected, bidirectional energy transfer is
observed: around $60 \%$ of total energy transfers upscale and is
mainly dissipated by the hypoviscosity, while the other $40 \%$
transfers downscale and is mainly dissipated by the
hyperviscosity; this corresponds to $R\approx0.6$.  In
Table~\ref{Table} the value of $\epsilon_\mathrm{u}$ is obtained
by calculating the amount of energy dissipated by the
hypoviscosity and the value of $\epsilon$ is calculated from the
white-noise forcing applied in the numerical simulation.

Now, the right panel of Figure \ref{fig_energy_trans_V} shows the
comparison of structure functions obtained in several different
ways.  The blue curve shows the structure function 
directly measured from the statistics of the velocity and
magnetic fields.  The black curve is the theoretical formula
(\ref{V_theo}) using the observed value of $\epsilon_\mathrm{u}$
as well as  the forcing
wavenumber $k_f=32$ and the total energy transfer $\epsilon$
known from the numerical setup.  The red curve is a
least-square fitting of the theoretical result (\ref{V_theo})
using only the four measured points from the blue curve marked in green
squares.  We choose these four points as a test because we need
to capture the sign transition of the third-order structure
function and we intentionally avoid choosing points in the
classic inertial ranges to distinguish our theory from the past
ones: the left three points are around the region of sign change
and the last point is around the forcing scale.  The parameters
used in the fittings are also shown in Table~\ref{Table}.  This
comparison shows that the fitting based on our global theory
using only four measured structure function values works well in
determining the bidirectional energy flux rate within a 5\%
error.
\begin{table}
	\centering
	\setlength{\tabcolsep}{0.5em} % for the horizontal padding
	{\renewcommand{\arraystretch}{1.5}
		\begin{tabular}{c | c c c}
			& $\epsilon$            & $R=\epsilon_u/\epsilon$ & $k_f$ \\ \hline 
			$V_{fit1}$ & $1.000\times 10^{-2}$ & $0.5845$                & $32.00$ \\
			$V_{fit2}$  & $0.958\times 10^{-2}$ & $0.5786$                & $32.06$ \\          
		\end{tabular}
	}
	\caption{Comparison of the coefficients of two fitting curves shown in the right panel of Figure \ref{fig_energy_trans_V}.}
	\label{Table}
\end{table}    

The right panel of Figure \ref{fig_energy_trans_V} also shows
that the dissipation at large-scale due to hypoviscosity brings
about a nonnegligible discrepancy between the theory
(\ref{V_theo}) in zero-viscosity limit and the numerical data.
To capture this large-scale dissipative correction we include the
hypoviscosity $\mc{L}_{Dl}=\alpha\nabla^{-2}$ but omit the
hyperviscosity effect in (\ref{V_damp1}) to obtain the viscous
expression of the third-order structure function
\begin{equation}\label{V_dampMHD}
\begin{aligned}
V_{d} &= -\frac{4}{r}\int_{0}^{r}
s\sbr{\alpha\nabla^{-2}\br{\ovl{\bu\cdot\bu'}(s)+\ovl{\bB\cdot\bB'}(s)}-\epsilon_\mathrm{u}}
\dd s + 2 \epsilon_\mathrm{u}r -
\frac{4\epsilon}{k_f}J_1(k_fr)&\\ 
&= - \frac{2\alpha}{r}\int_{0}^{r} s\br{\ovl{\delta\psi^2}(s)+\ovl{\delta A^2}(s)} \dd s + 2 \epsilon_\mathrm{u}r - \frac{4\epsilon}{k_f}J_1(k_fr),&
\end{aligned}
\end{equation}
where $\psi$ and $A$ are the stream functions for $\bu$ and
$\bB$, respectively, and we have used the identity
$\nabla^2\ovl{AA'} = -\ovl{\nabla A\cdot\nabla'A'}$, which holds
for arbitrary scalar fields $A$ with isotropic statistics.  The excellent
match between (\ref{V_dampMHD}) and the numerical
data verifies the validity of (\ref{V_damp1}).  Thus, if the
damping form is known and the corresponding second-order
structure function can be measured, we can make use of them to fit
the data in a broader range to detect energy transfer.

%\begin{figure}
%	\centering
%	
%	\caption{}
%	\label{fig_V}
%\end{figure}

\section{Discussion}

To test our global results (\ref{V_theo}) and (\ref{V_damp1}), we
deliberately used a relatively low-resolution 2D MHD simulation,
which provides imperfect inertial ranges.  This severely limits
the applicability of classic local theories but not of the new
global theory.  Indeed, due to the limited resolution, the direct
numerical data (blue) shows that the energy inertial ranges which
have $V\sim r$ behavior are not observed. Similarly, because of
the non-negligible influence from the forcing scale, a straight
line corresponding to $V\sim r^3$ is also not clear. These make
the traditional process based on classic local theories of
fitting straight lines in a log-log plot to obtain the
information of energy flux impossible, but our global theory can
achieve it.  In addition, since our global theory only contains
three parameters and applies to a broader range containing
forcing scale, we can make use of more data information and
thereby detect the forcing scale.

The sublimits of our global expression (\ref{V_theo}) match those
of the classic inertial-range results (cf. (\ref{V_expan}))
implying that our theory captures the transitions of inertial
ranges.  Also, it implies that simply ``gluing" the theories of
different inertial ranges for turbulence with unidirectional
energy transfer to obtain a global theory is fallacious, because
the constant in front of the $r^3$ depends on the total energy
input instead of the upscale transferred energy alone.  Also,
this expansion brings about a new perspective to understanding
the ``enstrophy" range. In \citet{Kraichnan1967}'s argument, the
simultaneous conservation of both energy and enstrophy results in
an upscale energy transfer and a downscale enstrophy transfer,
and correspondingly in the enstrophy inertial range the
third-order structure function has an $r^3$ dependence.  However,
our theory shows that as long as there exists nonzero upscale
energy transfer, an $r^3$ dependence of third-order structure
function exists as a natural consequence of asymptotic expansion,
but the presence of a constant downscale ``enstrophy" flux is not
necessary with the ``enstrophy" a preserved quantity without
external force and dissipation, which is the case for 2D MHD
turbulence.

In this paper, we present a general framework for the
inertial-range third-order structure-function global theory that
captures  bidirectional energy transfer and resolves the
forcing scale in homogeneous isotropic turbulence.  This theory
has three parameters, $\epsilon_\mathrm{u}$,
$\epsilon_\mathrm{d}$ and $l_f$, that  describe the upscale energy
flux magnitude, downscale energy flux magnitude and the forcing
scale.  The classic local theories that are applicable away from
the forcing scales are recovered as sublimits of this global
theory, which captures the transitions as well.

In the present theory we assumed that the energy input is
$\delta$-centered at one wavenumber $k_f$, but considering that
when assuming a $\delta$-centered external forcing we are solving
a Green's function for equation (\ref{abs_eq_stea}) we can
express the expression of third-order structure function with a
general distribution of energy input rate after a convolution.
Thus, our theory can be used to detect the unknown distribution
of energy input for a 2D turbulent system.

%We have mainly focused on the relation between the third-order structure function and spectral energy transfer in an idealized the zero viscosity limit where the dissipations at the large and scales are assumed to happen at wavenumber $k=0$ and $k=\infty$, respectively. 
%However, this assumption is not justified, and we know that large-scale damping influences the behaviors of both large and small scales in various turbulence systems. 
%E.g., in 2D turbulence the linear Ekman damping introduces the log-correlation to the energy spectrum at the enstrophy inertial range \citep{Kraichnan1971}, and it bounds the range of inverse energy cascade \cite[e.g.][]{SmithKS2002}.
%We do not want to introduce a closure that links second- and third-order structure functions to calculate the shape of them, instead, we propose the relation
%\begin{equation}\label{V_damp}
%V - \frac{2}{r}\int_{0}^{r} s\mc{L}_DC(s) \dd s = 2 \epsilon_\mathrm{u}r - \frac{4\epsilon}{k_f}J_1(k_fr),
%\end{equation}
%that extends the relation  (\ref{V_theo}) of zero-viscosity limit to include the effect of damping effect at finite scales.
%Here, the dissipation operator $\mc{L}_D$ is so defined that $D=\mc{L}_DC$ in (\ref{abs_eq}). 
%In the 2D turbulence, $\mc{L}_D=-\alpha + \nu\nabla^2$ corresponding to (\ref{2d_damp}).
%Relation (\ref{V_damp}) is shown in a simplified form with the linear damping $\mc{L}_D=-\alpha$ in (6.2) of \cite{Xie2018}.
%%In the numerical simulation of MHD turbulence shown in \S \ref{sec_MHD}, 

As to the finite damping effect, it is shown in \cite{Xie2018} that for 2D turbulence with damping operator $\mc{L}_D=-\alpha+\nu\nabla^2$ the damping effect in (\ref{V_damp1}) tends to zero as $\alpha$ and $\nu$ tends to zero. But the comparable smallness of the damping effect in (\ref{V_damp1}) remains to be studied carefully in other turbulence system. And it is important to justify that the different operations to the large- and small-scale damping effects is general and therefore in the limit of zero viscosity large-scale damping impacts the third-order structure function at the leading order while the influence of small-scale damping is negligible. 
 
In the main text of this paper we only show the 2D theory for
the reason that we can test it numerically. We close the paper
by present the third-order structure function expression
analogous to \eqref{V_theo} for 1D (Burgers)
and 3D isotropic turbulence with bidirectional energy transfer:
\begin{equation}\label{1D_V}
\begin{aligned}
\textrm{1D:} \quad V&= 4\epsilon_\mathrm{u}r - 4 \epsilon \frac{\sin\br{k_fr}}{k_f}\\
& = \left\{\begin{matrix}
-4\epsilon_\mathrm{d} r + \dfrac{2}{3}\epsilon k_f^2r^3 + O\br{(k_fr)^5}, \quad  \br{k_fr\ll 1}, \\
4\epsilon_u r+ O\br{1}, \quad \br{k_fr\gg 1}.
\end{matrix}\right.
\end{aligned}
\end{equation}
\begin{equation}\label{3D_V}
\begin{aligned}
\textrm{3D:} \quad V&= \frac{4}{3}\epsilon_\mathrm{u}r - 4 \epsilon \frac{\sin\br{k_fr}-Kr\cos\br{k_fr}}{k_f^3r^2}\\
& = \left\{\begin{matrix}
-\dfrac{4}{3}\epsilon_\mathrm{d} r + \dfrac{2}{15}\epsilon k_f^2r^3 + O\br{(k_fr)^5}, \quad  \br{k_fr\ll 1}, \\
\dfrac{4}{3}\epsilon_u r+ O\br{\br{k_fr}^{-1}}, \quad \br{k_fr\gg 1}.
\end{matrix}\right.
\end{aligned}
\end{equation}
Note that the 3D result for small $k_fr$ gives $V = -\frac{4}{3}\epsilon_\mathrm{d}r$. For classic 3D turbulence, considering the relation between $V$ and the longitudinal third-order structure function
\begin{equation}
V = \ovl{\delta u_L^3} + \frac{1}{3}\frac{\dd }{\dd r}(r\ovl{\delta u_L^3})
\end{equation}
we recover the $-4/5$ law of \citet{Kolmogorov1941}'s theory: $\ovl{\delta u_L^3}=-\frac{4}{5}\epsilon_\mathrm{d}r$. 
The detailed derivations of the 1D and 3D expression are shown in \S \ref{sec_1D3D}.

\vspace{1em}
We grateful to Andrew Majda and Shafer Smith for discussions that help to improve this paper.
We gratefully acknowledge financial support from the United States National Science Foundation grant DMS-1312159 and Office of Naval Research grant N00014-15-1-2355.

\appendix

\section{Derivation of the expressions of third-order structure functions in 1D and 3D}\label{sec_1D3D}

The key for deriving the expressions of third-order structure functions is obtaining the relation between structure function and energy flux, e.g. (\ref{V_F}) in the main text, then we substitute the idealized energy flux function (cf. (\ref{F_heavi}) in the main text) to obtain the final result. In this section we apply this procedure to 1D and 3D turbulence to obtain the expressions (\ref{1D_V}) and (\ref{3D_V}) in the main text.

\subsection{Third-order structure function in 1D}

In one dimension, the spectral energy flux (\ref{F_1}) can be expressed as
\begin{equation}\tag{S1}
F(K) = -\frac{1}{2\pi}\int_{-K}^{K} \dd k \int_{-\infty}^{\infty} \dd s \frac{1}{4} \frac{\dd V}{\dd s} \ex^{\ii ks},
\end{equation}
therefore
\begin{equation}\tag{S2}
\frac{\dd F}{\dd K} = -\frac{1}{4\pi} \int_{-\infty}^{\infty} \dd s\frac{\dd V}{\dd s} \ex^{\ii Ks}. \label{S2}
\end{equation}

Taking inverse Fourier transform to (\ref{S2}) we obtain
\begin{equation}\tag{S3}
\frac{\dd V}{\dd r} = -2 \int_{-\infty}^{\infty} \dd K \frac{\dd F}{\dd K} \ex^{-\ii Kr}, 
\end{equation}
so 
\begin{equation}\tag{S4}\label{S4}
\begin{aligned}
V = 4 \int_{0}^{\infty} \dd K F \frac{Kr\cos(Kr)-\sin(Kr)}{K^2},
\end{aligned}
\end{equation}
which express $V$ as a functional of $F$. 
Thus, substituting the idealized expression of bidirectional energy flux  
\begin{equation}\tag{S5}\label{S5}
F=-\epsilon_\mathrm{u} + (\epsilon_\mathrm{u}+\epsilon_\mathrm{d}) H(K-k_f) 
\end{equation}
into (\ref{S4}) we obtain
\begin{equation}
V= 4\epsilon_\mathrm{u}r - 4 \epsilon \frac{\sin\br{k_fr}}{k_f},
\end{equation}
which is (\ref{1D_V}).

\subsection{Third-order structure function in 3D}

In three dimension, the spectral energy flux (\ref{F_1}) can be expressed as
\begin{equation}\tag{S6}
F(K) = -\frac{1}{(2\pi)^3}\int_{0}^{K} \dd k \int_{0}^{\pi}\dd \phi \int_{0}^{\pi} \dd \theta k^2 \sin(\phi)
\int_{-\infty}^{\infty}\int_{-\infty}^{\infty}\int_{-\infty}^{\infty} \dd \bor \frac{1}{4} \nabla\cdot\bV \ex^{\ii \bk\cdot \bor},
\end{equation}
therefore under the assumption of isotropy, taking $K$-derivative and inverse Fourier transform we obtain
\begin{equation}\tag{S7}\label{S7}
\frac{1}{r^2}\frac{\dd }{\dd r}\br{r^2V} = -4 \int_{0}^{\infty} \dd K \frac{\dd F}{\dd K} \frac{\sin(Kr)}{Kr}.  
\end{equation}
So we can integrate (\ref{S7}) to obtain
\begin{equation}\tag{S8}\label{S8}
V = 4 \int_{0}^{\infty} \dd K F \frac{2Kr-3\sin\br{Kr}+Kr\cos\br{Kr}}{K^4r^3}.
\end{equation}
Thus, substituting the idealized expression of bidirectional energy flux (\ref{S5})
into (\ref{S8}) we obtain
\begin{equation}
V= \frac{4}{3}\epsilon_\mathrm{u}r - 4 \epsilon \frac{\sin\br{k_fr}-Kr\cos\br{k_fr}}{k_f^3r^2},
\end{equation}
which is (\ref{3D_V}).

\bibliographystyle{jfm}
\bibliography{T2Dref}

\end{document}